\documentclass{sig-alternate-05-2015}

\usepackage{booktabs}
\usepackage{hyperref}
\usepackage{paralist}
\usepackage{amsmath}
\usepackage{tikz}
\usepackage[flushleft]{threeparttable}
\usetikzlibrary{arrows, shapes}

\usepackage{graphicx}
\graphicspath{{./}}

\usepackage{color}

\begin{document}
\setlength{\belowdisplayskip}{5pt} \setlength{\belowdisplayshortskip}{5pt}
\setlength{\abovedisplayskip}{5pt} \setlength{\abovedisplayshortskip}{5pt}
\setlength{\textfloatsep}{6pt}
\setlength{\intextsep}{5pt}
% Copyright
%\setcopyright{acmcopyright}
%\setcopyright{acmlicensed}
%\setcopyright{rightsretained}
%\setcopyright{usgov}
%\setcopyright{usgovmixed}
%\setcopyright{cagov}
%\setcopyright{cagovmixed}

% DOI
%\doi{10.475/123_4}

% ISBN
%\isbn{123-4567-24-567/08/06}

%Conference
%\conferenceinfo{PLDI '13}{June 16--19, 2013, Seattle, WA, USA}

%\acmPrice{\$15.00}

%
% --- Author Metadata here ---
%\conferenceinfo{WOODSTOCK}{'97 El Paso, Texas USA}
%\CopyrightYear{2007} % Allows default copyright year (20XX) to be over-ridden - IF NEED BE.
%\crdata{0-12345-67-8/90/01}  % Allows default copyright data (0-89791-88-6/97/05) to be over-ridden - IF NEED BE.
% --- End of Author Metadata ---

\title{Classifying Developers into Core and Peripheral:\\An Empirical Study on Count and Network Metrics}
%\subtitle{[Extended Abstract]
%\titlenote{A full version of this paper is available as
%\textit{Author's Guide to Preparing ACM SIG Proceedings Using
%\LaTeX$2_\epsilon$\ and BibTeX} at

\numberofauthors{3} %  in this sample file, there are a *total*
\author{
% 1st. author
\alignauthor
Mitchell Joblin\\
       \affaddr{Siemens AG}\\
       \affaddr{Erlangen, Germany}\\
%       \email{trovato@corporation.com}
% 2nd. author
\alignauthor
Sven Apel, Claus Hunsen\\
       \affaddr{University of Passau}\\
       \affaddr{Passau, Germany}\\
%       \email{webmaster@marysville-ohio.com}
% 3rd. author
\alignauthor 
Wolfgang Mauerer\\
       \affaddr{Siemens AG\\OTH Regensburg}\\
       \affaddr{Munich/Regensburg, Germany}
}
% There's nothing stopping you putting the seventh, eighth, etc.
% author on the opening page (as the 'third row') but we ask,
% for aesthetic reasons that you place these 'additional authors'
% in the \additional authors block, viz.
%\additionalauthors{Additional authors: John Smith (The Th{\o}rv{\"a}ld Group,
%email: {\texttt{jsmith@affiliation.org}}) and Julius P.~Kumquat
%(The Kumquat Consortium, email: {\texttt{jpkumquat@consortium.net}}).}
% Just remember to make sure that the TOTAL number of authors
% is the number that will appear on the first page PLUS the
% number that will appear in the \additionalauthors section.

\maketitle
\begin{abstract}
Knowledge about the roles developers play in a software project is crucial to understanding the project's collaborative dynamics. Developers are often classified according to the dichotomy of core and peripheral roles. Typically, operationalizations based on simple counts of developer activities (e.g., number of commits) are used for this purpose, but there is concern regarding their validity and ability to elicit meaningful insights. To shed light on this issue, we investigate whether commonly used operationalizations of core--peripheral roles produce consistent results, and we validate them with respect to developers' perceptions by surveying 166 developers. Improving over the state of the art, we propose a relational perspective on developer roles, using developer networks to model the organizational structure, and by examining core--peripheral roles in terms of developers' positions and stability within the organizational structure. In a study of 10 substantial open-source projects, we found that the existing and our proposed core--peripheral operationalizations are largely consistent and valid. Furthermore, we demonstrate that a relational perspective can reveal further meaningful insights, such as that core developers exhibit high positional stability, upper positions in the hierarchy, and high levels of coordination with other core developers.
\end{abstract}

%\begin{CCSXML}
%<ccs2012>
%  <concept_id>10010520.10010553.10010562</concept_id>
%  <concept_desc>Computer systems organization~Embedded systems</concept_desc>
%  <concept_significance>500</concept_significance>
%  <concept_id>10010520.10010575.10010755</concept_id>
%  <concept_desc>Computer systems organization~Redundancy</concept_desc>
%  <concept_significance>300</concept_significance>
%  <concept_id>10010520.10010553.10010554</concept_id>
%  <concept_desc>Computer systems organization~Robotics</concept_desc>
%  <concept_significance>100</concept_significance>
%  <concept_id>10003033.10003083.10003095</concept_id>
%  <concept_desc>Networks~Network reliability</concept_desc>
%  <concept_significance>100</concept_significance>
%</ccs2012>  
%\end{CCSXML}
%
%\ccsdesc[500]{Computer systems organization~Embedded systems}


%
%  Use this command to print the description
%
%\printccsdesc

% We no longer use \terms command

%\keywords{ACM proceedings; \LaTeX; text tagging}

\section{Introduction}
%In open-source software development, there are numerous roles that contributors adopt, each with distinct characteristics and responsibilities. 
The popular ``onion'' model---first proposed by Nakakoji et al.~\cite{Nakakoji2002}---comprises eight roles typically appearing in open-source software projects. These roles extend from passive users of the software, to testers and, active developers. According to this model, there is a clear and intentional expression of the substantial difference in scale between the group sizes fulfilling each role. 
%That is to say, two groups fulfilling different roles will often differ in size by a significant margin. 
Multiple empirical studies gathered evidence of this model in terms of the heavy-tailed distribution describing the number code contributions per developer, which implies that a small fraction of developers is responsible for performing the majority of work~\cite{Mockus2002,Crowston2006}. 
%The form of this distribution also implies that most developers contributing to a project make only few or irregular contributions. 
From this simple observation, the distinction between different roles of developers is often coarsely represented as a dichotomy comprised of core and peripheral developers~\cite{Crowston2006}. In an abstract sense, \emph{core} developers play an essential role in developing the system architecture and forming the general leadership structure, and they have substantial, long-term involvement~\cite{Crowston2006}. In contrast, \emph{peripheral} developers are typically involved in bug fixes/small enhancements, and they have irregular or short-term involvement~\cite{Crowston2006}. 

%In this light, it is rather curious that open-source development results in any success at all, considering the fact that the vast majority of contributors (typically 80\%) are volatile in nature and have very limited engagement.

At first glance, it seems that the larger group of peripheral developers represents an unnecessary threat to project success, as their volatile nature results in the known problems of knowledge loss and inadequate changes~\cite{Terceiro10}. However, there is evidence that supports an alternative story: peripheral developers are just as critical to the project's success as core developers~\cite{Raymond1999}. Without the peripheral group, there is limited opportunity for a vetting process to identify and promote appropriate developers~\cite{Jensen2007}.
%This process is critical to establishing informed decisions regarding which developers are appropriate candidates for core positions \TODO{add citation}. If a project wishes to remain robust to developer turnover and achieve sustainable growth, then there must be an adequate talent pool from which to draw new core developers. 
Furthermore, peripheral developers are crucial to the ``many eyes'' hypothesis---which posits that all bugs become shallow when the source code is scrutinized by a sufficiently large number of people---that is often referenced as an explanation for why open-source development will inevitable result in a high-quality product~\cite{Raymond1999}. 
%Since core developers are in short supply, peripheral developers are the key for the project to benefiting from the consequences of many eyes.

Despite an understanding of the characteristics of core and peripheral developers and recognizing the importance of the interplay between these roles, there remain two open issues. Firstly, an appropriate core--peripheral operationalization is crucial for testing empirical evidence of proposed theories regarding collaborative aspects of software development. While several basic operationalizations have been proposed and loosely justified by abstract notions, they may be overly simplistic. For example, one common approach is to apply thresholding on the number of lines of code contributed by each developer~\cite{Mockus2002}, but this could result in incorrectly classifying developers making large numbers of trivial cleanups. 
%For example, the relationship between operationalizations based on technical contributions and those based on social contributions is unclear. 
%Secondly, the existing operationalizations are plausible, but (possibly overly) simple. For example, one common approach is to apply thresholding on the number of lines of code contributed by each developer, but this could result in incorrectly classifying developers making large numbers of trivial cleanups~\cite{Mockus2002}.Thirdly, it may be the case that developers do not naturally fit into two mutually exclusive groups and are more appropriately expressed according to a position on a spectrum where core and peripheral form the end points. In other words, the ordinal notion of \emph{coreness} is virtually unexplored. In all known cases, ordinal or ratio variables are used with a scientifically unvalidated threshold to establish the core--peripheral dichotomy~\cite{Crowston2006}. 
The second open issue arises from the fact that core--peripheral operationalizations are fundamentally based on simple counts (e.g., lines of code, number of commits, number of e-mails sent) that lack richness in describing the roles and that provide only limited insights into the possibly complex and global relationships between developers. Essentially, a relational perspective is missing. This prevents us from answering important questions such as: Is a certain relational pattern responsible for quality problems?

The contributions of this work can be summarized by two main achievements. Firstly, we statistically evaluate the agreement between the most commonly used operationalizations of core and peripheral developers by examining data stored in the version-control systems and developer mailing lists of 10 substantial open-source projects. We also performed a survey among 166 developers to establish a ground-truth classification of developer roles. The primary objective was to test whether existing operationalizations are consistent with respect to each other, and valid with respect to developer perception.
%As all operationalizations claim to measure the same high-level concept, they should be consistent with each other. 
Secondly, we establish and evaluate richer notions of developer roles with a basis in relational abstraction. More specifically, we adopt a network-analytic perspective to explore manifestations of core and peripheral characteristics in the evolving organizational structure of software projects, as operationalized by developer networks~\cite{joblin2015,joblin2015Ev}. Our conjecture is that, if the abstract characteristics of core and peripheral developers proposed in the literature are accurate, these roles should also manifest in ways that transcend simple counts of developer contributions. In particular, we explore stability patterns and structural embeddings of core and peripheral developers in the global organizational structure of a project, which contains more actionable information regarding organizational or collaborative issues than just a count of code contributions. 
%For example, in our study of 10 open-source projects, we identified characteristics of core and peripheral developers in our relational abstraction as manifestations in their positions and stability within the organizational structure. 
Most notable, we found in our study that core developers, in comparison to peripheral developers, exhibit significantly higher positional stability, exhibit higher global centrality in the organizational structure, and are arranged according to a relatively strict hierarchy. Furthermore, core developers are most likely to coordinate with other core developers, while peripheral developers are also most likely to coordinate with core developers. The implication is that peripheral developers cannot just be considered less active versions of core developers, but instead they represent an organizationally distinctive group that requires extensive support from developers in core positions, which has implications for the development of novel coordination tools and processes.

In summary, we make the following contributions:
\begin{compactitem}
\item
We statistically evaluate the agreement between the \emph{dichotomous} classifications of core and peripheral developers generated from  commonly used operationali\-zations---henceforth called \emph{count-based} operationaliza\-tions---by studying 10 substantial open-source projects, over at least one year of development, with data from two sources (version-control system and mailing list).
%
%\item
%We examine developers' roles as an ordinal variable by statistically evaluating the rank-based agreement between the \emph{continuous} metrics of developer \emph{coreness} that are used to classify developers as core or peripheral in the count-based operationalizations.
%
\item
We conduct a survey among 166 developers to establish a ground truth, which we used to judge the operationalizations regarding developer perception. 
\item
We identify features in the organizational structure and stability patterns of developers that plausibly capture the abstract notions of core and peripheral developers using network-analysis techniques, referred to as the \emph{network-based} operationalizations.
\item
We demonstrate that the developer classifications produced by the network-based operationalizations largely agree with the existing count-based operationalizations. Based on the developer survey results, we provide evidence that the network-based operationalizations are a better reflection of developer perception than the count-based operationalizations.
\item
We highlight and discuss a number of insights from our network-based operationalizations that are incapable of being provided by count-based operationalizations.
\end{compactitem}

All experimental data and source code are available at a supplementary Web site.\footnote{\url{http://siemens.github.io/codeface/fse2016}}

\section{Background \& Related Work} \label{sec:related_work}
Previous research on core and peripheral developers has established an understanding of the characteristics possessed by each group. Researchers have examined the roles from two distinct perspectives: from a social perspective, by studying communication and collaboration patterns~\cite{Cataldo2008,Manteli2014,Mockus2002}, and from a technical perspective, by studying patterns of contributions of developers to technical artifacts~\cite{Crowston2006,Mockus2002,Terceiro10,Jensen2007}. Regarding social characteristics, core developers play a central role in the communication and leadership structure~\cite{Cataldo2008} and have substantial communication ties to other core developers, especially in projects with a small developer community (10--15 people)~\cite{Manteli2014,Mockus2002}. Regarding technical characteristics, core developers typically exhibit strong ownership over particular files that they manage, they often have detailed knowledge of the system architecture, and they have demonstrated themselves to be extremely competent~\cite{Crowston2006,Mockus2002,Terceiro10,Jensen2007}. In contrast, peripheral developers are primarily involved in identifying code-quality issues and in proposing fixes, while also participating moderately in development-related discussions~\cite{Mockus2002}. Since the roles of developers are not static, prior research has also investigated temporal characteristics of core and peripheral developers in terms of the advancement process to achieving core-developer status. Advancement is typically merit-based and often involves long-term, consistent, and intensive involvement in a project~\cite{Mockus2002,Crowston05,Ye2003,Jensen2007}.

Many of the aforementioned studies applied empirical methods based on interviews, questionnaires, personal experience reports, and manual inspections of data archives to identify characteristics of core and peripheral developers. An alternative line of research has attempted to operationalize core and peripheral developers using data available in software repositories, such as version-control systems~\cite{Mockus2002,Terceiro10,Robles2009,Robles2006,Oliva2012,deSouza2005}, bug trackers~\cite{Crowston2006}, and mailing lists~\cite{Oliva2012,Bird2007}. By operationalizing the notion of core and peripheral developers, these studies have taken important steps towards gaining insight that is not attainable with (more) manual approaches, including evaluating and basing conclusions on results from hundreds of projects~\cite{Crowston05}. 

Despite the existence of numerous operationalizations, we have very limited knowledge about their validity, though. There is a reasonable cause for concern that some corresponding metrics are overly simple: Most operationalizations are single-dimension values that represent the developer's activity level (e.g., the number of commits made), with a corresponding threshold based on a prescribed quantile. A commonly used approach is to count the number of commits made by each developer, and then to compute a threshold at the 80\% percentile. Developers that have a commit count above the threshold are considered core, developers below are considered peripheral~\cite{Crowston2006,Mockus2002,Terceiro10,Robles2009,Robles2006}. This threshold was rationalized by observing that the number of commits made by developers typically follows a Zipf distribution (which implies that the top 20\% of contributors are responsible for 80\% of the contributions)~\cite{Crowston2006}. Mockus et. al similarly found empirical evidence in Mozilla browser and Apache web server that a small number of developers are responsible for approximately 80\% of the code modifications~\cite{Mockus2002}. Further attempts have been made to investigate the difference between core and peripheral developers by using basic social-network centrality metrics and a corresponding threshold~\cite{Oliva2012,deSouza2005,Bird2007}. In these cases, developer networks have been constructed on a dyadic domain of either mutual contributions to mailing-list threads or source-code files.

\section{count-based\\Operationalizations} \label{sec:existing_metrics}
Based on a review of the existing literature, we have identified three variations of count-based operationalizations of core and peripheral roles~\cite{Mockus2002,Terceiro10,Robles2009,Robles2006,Oliva2012,deSouza2005,Crowston2006,Bird2007}. In these studies, metrics are used with a corresponding threshold to define a dichotomy composed of core and peripheral developers. We apply the standard 80th percentile threshold, because of its wide use and its justification based on the data following a Zipf distribution (see Section~\ref{sec:related_work}). Two operationalizations capture technical contributions to the version-control system and one captures social contributions to the developer mailing list.

\textbf{\emph{Commit count}}
is the number of commits a developer has authored (merged to the master branch). A commit represents a single unit of effort for making a logically related set of changes to the source code. Core developers typically make frequent contributions to the code base and should, in theory, achieve a higher commit count than peripheral developers. 

\textbf{\emph{Lines of code (LOC) count}}
is the sum of added and deleted lines of code a developer has authored (merged to the master branch). Counting LOC, as it relates to developer roles, follows a similar rationale to commit count. As core developers are responsible for the majority of changes, they should reach higher LOC counts than peripheral developers. A potential source of error is that developers writing inefficient code or changing a large number of lines with trivial alterations (e.g., whitespace changes) could artificially affect the classification.

\textbf{\emph{Mail count}}
is the number of mails a developer contributed to the developer mailing list. Core developers often posses in-depth technical knowledge, and the mailing list is the primary public venue for this knowledge to be exchanged with others. Core developers offer their expertise in the form of: making recommendations for changes, discussing potential integration challenges, or providing comments on proposed changes from other developers. Typically, peripheral developers ask questions or ask for reviews on patches they propose. Core developers often participate more intensively and consistently and have greater responsibilities than peripheral developers, in general. This should result in core developers making a large number of contributions to the mailing list. This is still only a very basic metric because a developer answering many questions and one asking many questions will appear to be equivalent, and there is no relational expression, so who is speaking with whom or with how many people is completely ignored.

\medskip
Each of the above metrics has a foundation rooted in our current empirical understanding of the characteristics of core and peripheral developers, but in the end, they are all relatively simple abstractions of a potentially multifaceted and complex concept. A comparison between the resulting classification of developers from these different metrics will provide valuable insights into whether systematic errors exist in these count-based operationalizations, which we perform in Section~\ref{sec:compare_existing}. However, the focus of these metrics is still only to assign developers exclusive membership to one of two unordered sets---without relational information between sets or within the sets---the insights offered by the classification are of limited practical value. To address this shortcoming, we propose a relational view on developer coordination and communication to extract insights that are of greater practical relevance to software engineering.

\section{A Network Perspective} \label{sec:methodology}
A developer network is a relational abstraction that represents developers as nodes and social or technical relationships between developers as edges. The promise of a network perspective is greater practical insights concerning the organizational and collaborative relationships between developers~\cite{Cataldo2008,Meneely2011,Cataldo2010,deSouza2005,Bird}. But to what extent can this promise be fulfilled? So far, we know that developer networks, when carefully constructed on version-control-system and mailing-list data, can be both accurate in reflecting developer perception and reveal important functional substructure, or communities, with related tasks and goals~\cite{joblin2015,Bird}. What can be elicited from developer networks regarding the core--peripheral dichotomy has not yet been greatly explored, and that is our intention in this work. Practical opportunities for network insights are, for example: Identifying core developers that are overwhelmed by the peripheral developers they need to coordinate with; structural equivalence (that is two nodes with the same neighbors) could reveal which core developers have similar knowledge or technical abilities, which helps to determine appropriate developers for sharing or shifting development tasks; structural holes between core developers may indicate deteriorating coordination; or a single globally central core developer may indicate an important organizational risk. 

\subsection{Network Model}
We now present the details of our network-analytic approach for analyzing data from the version-control systems and mailing lists to examine relational characteristics of core and peripheral developers. Intuition and prior research lead us to the conclusion that the role a developer fulfills can change over time~\cite{Jensen2007}. For this reason, we analyze multiple contiguous periods over one year of a project in question using overlapping analysis windows. Each analysis window is three months in length, and each subsequent analysis period is separated by two weeks~\cite{joblin2015Ev}. We chose three-month analysis windows, because it has been shown that, beyond this window size, the development community does not change significantly, but temporal resolution in their activities is lost~\cite{Meneely2011}.

\paragraph*{Social-network abstraction}
For a given project, we download the mailing lists archives either from \emph{gmane} using \emph{nntp-pull} or directly from the project's homepage to obtain an \emph{mbox} formatted file containing all messages sent to the mailing list. Most projects have different mailing lists for different purposes. We consider only the primary mailing list for development-related discussions. We apply several preprocessing steps to remove duplicated messages, normalize author names, and organize the mails into threads using the \emph{Message-IDs} and \emph{In-Reply-To-IDs}~\cite{Feinerer2008}. Furthermore, we decompose the \emph{From} lines of each mail into a (name, e-mail address) pair. In some cases, only an e-mail address or only a name is possible to recover, and this can present issues with identifying all mails that a single person sent. To resolve multiple aliases to a single identity, we use a basic heuristic approach similar to the one proposed by Bird et al.~\cite{Bird2006}. Despite the potential problems regarding author-name resolution---as developers accumulate valuable credibility through contributions to the mailing list---it is counterproductive for highly active individuals to use multiple aliases and conceal their identity. To construct a network representation of developer communication, we apply the standard approach, where edges are added between individuals that make subsequent contributes to a common thread of communication~\cite{Bird2006}.

\paragraph*{Technical-network abstraction}
Data in version-control systems are organized in a tree structure composed of commits. We analyze only the main branch of development, as a linearized history, by flattening all branches merged to master. Furthermore, we analyze only the \emph{authors} of commits, not the committer (which are expressed differently for Git), and attribute the commit to a unique individual using the same aliasing algorithm as for the mailing-list data. We count lines of code for each commit based on diff information, where the total line count is the sum of added and deleted lines. The network representation of developer activities in the version-control system is constructed using fine-grained \emph{function}-level information, which was observed to produce authentic networks that agree with developer perception~\cite{joblin2015}. In this approach, source-code structure is used to identify when two developers edit related lines of code. We enhance the network with semantic-coupling relationships between functions, which has shown to also reflect developer perception of artifact coupling~\cite{bavota2013}. The semantic relationships are identified by making use of the domain-specific words that are embedded in the textual content of the implementation (e.g., variable and function identifiers)~\cite{joblin2015Ev}. The end result is a relational abstraction that expresses links between developers contributing technically related changes, which signify the existence of task interdependencies between the developers.

\subsection{Core and Peripheral Developers in\\Developer Networks} \label{sec:proposed_metrics}
In Section~\ref{sec:related_work}, we noted that we should expect manifestations of the distinct qualities of core and peripheral developers in ways that transcend the count-based operationalizations introduced in Section~\ref{sec:existing_metrics}---an expectation that is also backed by a survey among 166 open-source developers (see Section~\ref{sec:support_network_perspective}). Next, we introduce five corresponding network-based operationalizations that rely on developer networks and their evolution.

\textbf{\emph{Degree centrality}}
aims at measuring local importance. It represents the number of ties (edges) a developer has to other developers~\cite{Brandes2005}. As essential members of the leadership/coordination structure, core developers associate with other core members and with peripheral developers that require their technical guidance. Peripheral developers are likely involved in only a small number of isolated changes and thus have only a limited number of interactions with other members of the development community. The expectation is that core developers then have a larger degree than peripheral developers.   

\textbf{\emph{Eigenvector centrality}}
is a global centrality metric that represents the expected importance of a developer by either connecting to many developers or by connecting to developers that are themselves in globally central positions~\cite{Brandes2005}. Since core developers are critical to the leadership and coordination structure, we expect them to occupy globally central positions in the developer network.

\textbf{\emph{Hierarchy}}
is present in networks that have nodes arranged in a layered structure, such that small cohesive groups are embedded within large, less cohesive groups. In a hierarchical network, nodes with high degree tend to have edges that span across cohesive groups, thereby lowering their clustering coefficient~\cite{ravasz2003hierarchical}. Prior work has shown that developers tend to form cohesive communities~\cite{joblin2015}, and we expect core developers to play a role in coordinating the effort of these communities of developers. If this is true, then core developers should have a high degree and low clustering coefficient, placing them in the upper region of the hierarchy, while peripheral developers should exhibit a comparatively low degree and high clustering coefficient, placing them in the lower region of the hierarchy. 

\textbf{\emph{Role stability}}
is a temporal property of how developers transition between roles. For this reason, we investigate the patterns of developers' transitions through different roles by observing changes in the corresponding developer network over time. As core developers typically attain their credibility through consistent involvement and often have accumulated knowledge in particular areas of the system over substantial time periods (see Section~\ref{sec:related_work}), we expect their stability in the developer network to be higher than for peripheral developers. We operationalize developer stability by estimating the probability that a developer in a given role transitions to another role. For each developer the role during each development window is determined using the degree-centrality operationalization. The time ordered sequence of roles for each developer is then used in a maximum-likelihood estimation to solve for each state transition parameter (e.g., the probability that a core developer transitions to a peripheral role)~\cite{bishop2006pattern}.

\textbf{\emph{Core--peripheral block model}}
is a formalization, proposed in the social-network literature, that captures the notion of core--periphery structure based on an adjacency-matrix representation. The block model specifies the core--core region of the matrix as a 1-block (i.e., completely connected), the core--peripheral regions as imperfect 1-blocks, and the peripheral--peripheral region as a 0-block~\cite{zhang15}. Intuitively, this model describes a network as a set of core nodes, with many edges linking each other, surrounded by a loosely connected set of peripheral nodes that have no edges connecting each other. Of course, this idealized block model is rarely observed in empirical data~\cite{Borgatti2000}. Still, we are able to draw practical consequences from this formalization by estimating the edge presence probability of each position to test if core and peripheral developers (operationalized by degree centrality) occupy core and peripheral network positions according to this block model. From the block model, one can mathematically reason that the probability of observing an edge in each block is distinct and related according to $p_{\text{core--core}} > p_{\text{core--periph}} > p_{\text{periph--periph}}$~\cite{zhang15}. This model aligns with empirical data that indicate that the core developers are typically well-coordinated and are expected to be densely connected in the developer network~\cite{Mockus2002}. Since peripheral developers often rely on the knowledge and support of core developers to complete their tasks, it follows that peripheral developers often coordinate with core developers, and only in rare cases would we expect substantial coordination between peripheral developers. This expected behavior aligns very well to the formalized notion of core--periphery positions from social-network analysis.

\section{Empirical Study}
We now present the details of our empirical study to test for agreement between the different count-based operationalizations of core and peripheral developer roles and to identify richer relational characteristics of these roles represented by our proposed network-based operationalizations.

\subsection{Subject Projects}
We selected ten open-source projects, listed in Table~\ref{table:oss_project_data}, to
study the core--peripheral developer roles. We specifically
chose a diverse set of projects to avoid biasing the results.
The projects vary by the following dimensions: (a) size (source lines of code from
50KLOC to over 16 MLOC, number of developers from 15
to 1000), (b) age (days since first commit), (c) technology
(programming language, libraries used), (d) application domain
(operating system, development, productivity, etc.), (e) development
process employed. Developers of the project referred to as Project X
have requested that their project name remain anonymous.

\begin{table*}[ht]
\centering
\caption{Overview of subject projects} 
\vspace{1ex}
\begin{threeparttable}
\small
\label{table:oss_project_data}
\begin{tabular}{lllrrrcrrrrrr}
  \toprule
  & & & & & & & \multicolumn{3}{c}{Edge Probabilities} & \multicolumn{2}{|c}{Hierarchy}\\
  \cmidrule{8-12} 
Project & Domain & Lang & Devs & SLOC & Commits & Date & C--C & C--P & P--P & Rho\tnote{1} & P value\\ 
  \midrule
  Project X & User & C/++, JS & 826 & 10M & 276K & 2015/12/05 & 9.75e-02 & 4.19e-03 & 2.70e-03 & -0.552 & 5.51e-33 \\ 
  Django & Devel & Python & 100 & 430K & 41K & 2015/12/06 & 2.95e-01 & 9.09e-03 & 3.08e-03 & -0.812 & 1.28e-06 \\ 
  FFmpeg & User & C & 103 & 1M & 78K & 2015/11/08 & 5.50e-01 & 2.44e-02 & 5.16e-03 & -0.725 & 7.10e-06 \\ 
  GCC & Devel & C/++ & 122 & 7.5M & 144K & 2015/11/03 & 4.07e-01 & 1.84e-02 & 1.01e-02 & -0.646 & 1.12e-04 \\ 
  Linux & OS & C & 1467 & 18M & 637K & 2015/12/05 & 2.39e-02 & 5.93e-04 & 3.60e-04 & -0.689 & 6.06e-62 \\ 
  LLVM & Devel & C/++ & 180 & 1.1M & 62K & 2015/11/02 & 7.80e-01 & 5.54e-02 & 2.62e-02 & -0.778 & 8.72e-24 \\ 
  PostgreSQL & Devel & C & 17 & 1M & 40K & 2015/12/05 & 1.00e+00 & 1.62e-01 & 5.13e-02 & -0.871 & 1.31e-03 \\ 
  QEMU & OS & C & 134 & 1M & 43K & 2015/11/02 & 3.20e-01 & 1.95e-02 & 1.16e-02 & -0.756 & 4.76e-07 \\ 
  U-Boot & Devel & C & 142 & 1.3M & 35K & 2015/11/01 & 2.00e-01 & 7.59e-03 & 4.20e-03 & -0.728 & 8.27e-05 \\ 
  Wine & User & C & 62 & 2.8M & 110K & 2015/11/06 & 3.46e-01 & 2.91e-02 & 1.28e-02 & -0.832 & 1.04e-05 \\ 
   \bottomrule
\end{tabular}
\begin{tablenotes}
\item[1] Spearman's correlation coefficient
\end{tablenotes}
\end{threeparttable}
\end{table*}

\subsection{Research Questions}
While many approaches exist to classify developers into core and peripheral, no substantial evidence has been accumulated to validate the consistency of these different operationalizations. Crowston et al.~\cite{Crowston2006} investigated three operationalizations of core and peripheral developers, but they focused only on bug-tracker data and neglected code authorship entirely. Olivia et al.~\cite{Oliva2012} dedicated attention on developing a more detailed characterization of so-called ``key developers'', which is similar to the core-developer concept. They investigated mailing lists and version-control systems with three operationalizations to classify developers as core or peripheral. Their results indicate that there is some evidence of agreement between the different operationalizations, but this was only shown for a single release of a single small project with only 16 developers, in total, and 4 core developers. We improve over the state of the art by considering a larger and more diverse set of projects with larger developer communities, and by analyzing, at least, one year of development, to evaluate the temporal stability of our results. 
%Furthermore, we evaluate the \emph{continuous} notion of coreness (see Section\ref{sec:meas_agree}), which has been noted in the literature as an unaddressed issue~\cite{Crowston2006}.
While each of the approaches for classifying core and peripheral developers is inspired by common abstract notions rooted in empirical results, it has not been shown that the approaches agree. It may be the case that they capture orthogonal dimensions of the same abstract concept, which gives rise to our first research question:
\vskip 1ex
\noindent\textbf{RQ1: Consistency}---\emph{Do the commonly applied operationalizations of core and peripheral developers based on version-control-system and mailing-list data agree with each other?}
\vskip 1ex
Compared to the extent of our knowledge regarding the characteristics of core and peripheral developers, existing count-based operationalizations are relatively simple. Since core developers often have strong ownership over particular files and play a central role in coordinating the work of others on those artifacts~\cite{Cataldo2008,Mockus2002}, we would expect core developers to differ, in a relational sense, from peripheral developers in how they are embedded in the communication and coordination structure. Furthermore, as core developers typically achieve their status through long-term and consistent involvement~\cite{Jensen2007}, we expect their temporal stability patterns to differ from peripheral developers.
\vskip 1ex
\noindent\textbf{RQ2: Positions \& Stability}---\emph{Do the differences between core and peripheral developers manifest in relational terms within the communication and coordination structure with respect to their positions and stability?}
\vskip 1ex
The utility offered by an operationalization is limited by the extent to which the operationalization is able to accurately capture a real-world phenomenon. So far, it is unclear to what extent the core--peripheral operationalizations reflect developer roles as seen by their peers. We explore whether relational abstraction, as in the network-based operationalizations, improves over the count-based operationalizations by more accurately reflecting developer perception through explicit modeling of developer-developer interactions.
\vskip 1ex
\noindent\textbf{RQ3: Developer Perception}---\emph{To what extent do the various count-based and network-based operationalizations agree with developer perception?}

\subsection{Hypotheses} \label{sec:hypotheses}
The existing count-based operationalizations of core and peripheral developers discussed in Section~\ref{sec:existing_metrics} claim to be valid measures, and if this is a matter of fact, we expect to reach consistent conclusions about which developers of a given project belong to the core group and which belong to the peripheral group. Due to finite random sampling and sources of noise, we expect imperfect agreement between two operationalizations even if they are consistent in capturing the same abstract concept. However, if the operationalizations are consistent, the level of agreement in the results should be significantly greater than the case of random assignment of developer roles (i.e., core or peripheral). Our null model for zero agreement is the amount of agreement that results from two operationalizations that assign classes according to a Bernoulli process.\footnote{A Bernoulli process generates a sequence of binary-valued random variables that are independent and identically distributed according to a Bernoulli distribution. The process is essentially simulating repeated coin flipping.} To operationalize agreement between two binary classifications (core or peripheral) of a given set of developers, we use Cohen's kappa:
\begin{equation}
\kappa = \frac{p_{o} - p{e}}{1 - p_{e}},
\end{equation}
\noindent
where $p_{o}$ is the number of times the two classifications agree on a role of a developer, divided by the total number of developers and where $p_{e}$ is the expected probability of agreement when there is random assignment of roles to developers, but the proportion of each class is maintained. Cohen's kappa is more robust than simple percent agreement because it incorporates the effect of agreement that occurs by chance~\cite{richard1977}. This characteristic is particularly important in our case since the frequency of roles is highly asymmetric as the majority of developers are peripheral and only a small fraction are core. The ranges for Cohen's kappa and corresponding strength of agreement are: 0.81--1.00 almost perfect, 0.61--0.80 substantial, 0.41--0.6 moderate, 0.21--0.40 fair, 0.00--0.20 slight, and $<0.00$ poor~\cite{richard1977}.

%Cohen's kappa is useful for determining how well dichotomous classifications of developers agree, however, we also adopt a finer-grained perspective and investigate the metrics agreement on an ordinal scale. To this end, we compare the metrics prior to applying the threshold at the 80th percentile. By comparing the metrics on an ordinal scale, we are able to avoid the influence of the chosen threshold on the results. We are interested in determining whether the various metrics agree their ranking of developers' coreness. 
%The absolute values are not of interest, rather we wish to know whether one metric ranks developers $d_{i} \geq d_{j}$ then if the second metric agrees, the inequality should hold for all $i$s and $j$s. 
%For this purpose, it is most appropriate to use standard rank-based correlation coefficient Spearman's rho.
%
\pagebreak
\vskip 1ex
\noindent\textbf{H1}---\emph{Existing count-based operationalizations of core and peripheral developers based on version-control-system and mailing\hspace{0mm}-list data are statistically consistent in classifying developer roles.}
\vskip 1ex
The abstract notion of core and peripheral developers discussed in Section~\ref{sec:related_work} emphasizes the multitude of ways the two groups differ (e.g., contribution patterns, knowledge, level of engagement, organization, responsibility, etc.). While existing operationalizations of core and peripheral developers are primarily based on simple metrics of counting high-level activities of developers, these metrics largely ignore the richness in the definition of core and peripheral roles. In particular, the dimension of time is largely ignored, though time plays a central role in the developer-advancement process~\cite{Jensen2007}. Likewise, the relative positions in the corresponding organizational structure are ignored. But a difference in how core and peripheral developers are embedded in the organizational structure is to be expected, since core developers have extensive involvement in the coordination of specific technical artifacts and preside over peripheral developers. Therefore, we expect to see manifestations of the differences between the two distinct groups of developers in the developer network.
\vskip 1ex
\noindent\textbf{H2}---\emph{The well-known abstract characteristics of core developers will manifest as distinct structural features in the corresponding developer network: Core developers will exhibit globally central positions, relatively high positional stability, and hierarchical embedding.}
\vskip 1ex
As core developers form the primary coordination structure, we expect to observe: many edges in the developer network between core developers, less edges between core and peripheral developers, and even fewer edges between peripheral developers. We investigate this hypothesis in terms of preferences between the groups to associate based on the probability of an edge occurring between them according to the core--peripheral block model (see Section~\ref{sec:proposed_metrics}).
\vskip 1ex
\noindent\textbf{H3}---\emph{Core developers will have a preference to coordinate with other core developers and peripheral developers will exhibit a preference to coordinate with core developers instead of other peripheral developers.}
\vskip 1ex
We expect developer networks to reveal core and peripheral developers, albeit in a more rich representation, with comparable precision to the currently accepted operationalizations. More specifically, we expect developer networks capture the core--peripheral property to an equally high standard as the currently accepted operationalizations; any disagreement should be on the order of the discrepancy between existing operationalizations.
\vskip 1ex
\noindent\textbf{H4}---\emph{The core--peripheral decomposition obtained from developer networks will be consistent with the core--peripheral decomposition obtained from the prior accepted operationalizations. The discrepancy in agreement will not exceed the amount observed between the existing operationalizations.}
\vskip 1ex
As the count-based operationalizations appear to reasonably capture simple aspects of developer roles, we expect a certain level of agreement between these operationalizations and developer perception. In the case of the network-based operationalizations, we expect even higher agreement with developer perception since the relational abstraction explicitly captures developer-developer interactions that are neglected by the count-based operationalizations.
\vskip 1ex
\noindent\textbf{H5}---\emph{The count-based operationalizations agree with developer perception but the network-based operationalizations exhibit higher agreement.}
\vskip 1ex

\subsection{Developer Perception} \label{sec:dev_survey}
To establish a ground-truth classification of developer roles, we designed an online survey in which we asked developers to report the roles of developer's in their project according to their perception. The goal of acquiring these data is to test whether the core--peripheral operationalizations are valid with regard to developer perception (not only to other operationalizations). A sample of the survey instrument can be found at the supplementary Web site.

We recruited participants for the study exclusively from the version-control-system data of our ten subject projects by identifying the e-mail addresses of individuals that made a commit within the three months prior to the survey date (see Table~\ref{table:oss_project_data}). This was to ensure that the selected developers have current knowledge of the project state, so that their answers are temporally consistent with our analysis time frame. One subject project, GCC, was excluded from the survey because the developer e-mail addresses are not available in the version-control system. For the remaining 9 projects, we sent recruitment e-mails to 3369 developers of which 166 elicited a complete response.

The survey includes two primary sections: The first section contains questions that require the developers to self-report their role in the project (core or peripheral) and to provide a textual description of the nature of their participation. This question is useful for identifying potential sampling-bias problems and to determine if developers' self-reported role is consistent with the answers provided by their peers. The second section includes a list of 12 developers, identified by name and e-mail address, sampled from their specific project. For each developer appearing in the list, the respondent was asked to provide a classification of the developer's role. Appropriate options are also available if the respondent did not know the developer in question or was unsure of the role. We applied the following sampling strategy to select the list of twelve developers: Five developers were randomly selected from the core group and five from the peripheral group, classified according the the commit count-based operationalization (see Section~\ref{sec:existing_metrics}). The remaining two developers were randomly selected from the direct neighbors, in the developer network, of the survey participant. We chose to use neighbors because it is likely that neighbors work directly together and would be aware of each other's roles.

\section{Results}
We now present the results of our empirical study and address the five hypotheses described in Section~\ref{sec:hypotheses}. For practical reasons, we are only able to present figures for a single project that is representative of the general results. Please refer to the supplementary Web site for the remaining project figures.

\subsection{Consistency of Count-Based\\Operationalizations} \label{sec:compare_existing}
\begin{figure}[t!]
\centering
\includegraphics[width=\linewidth]{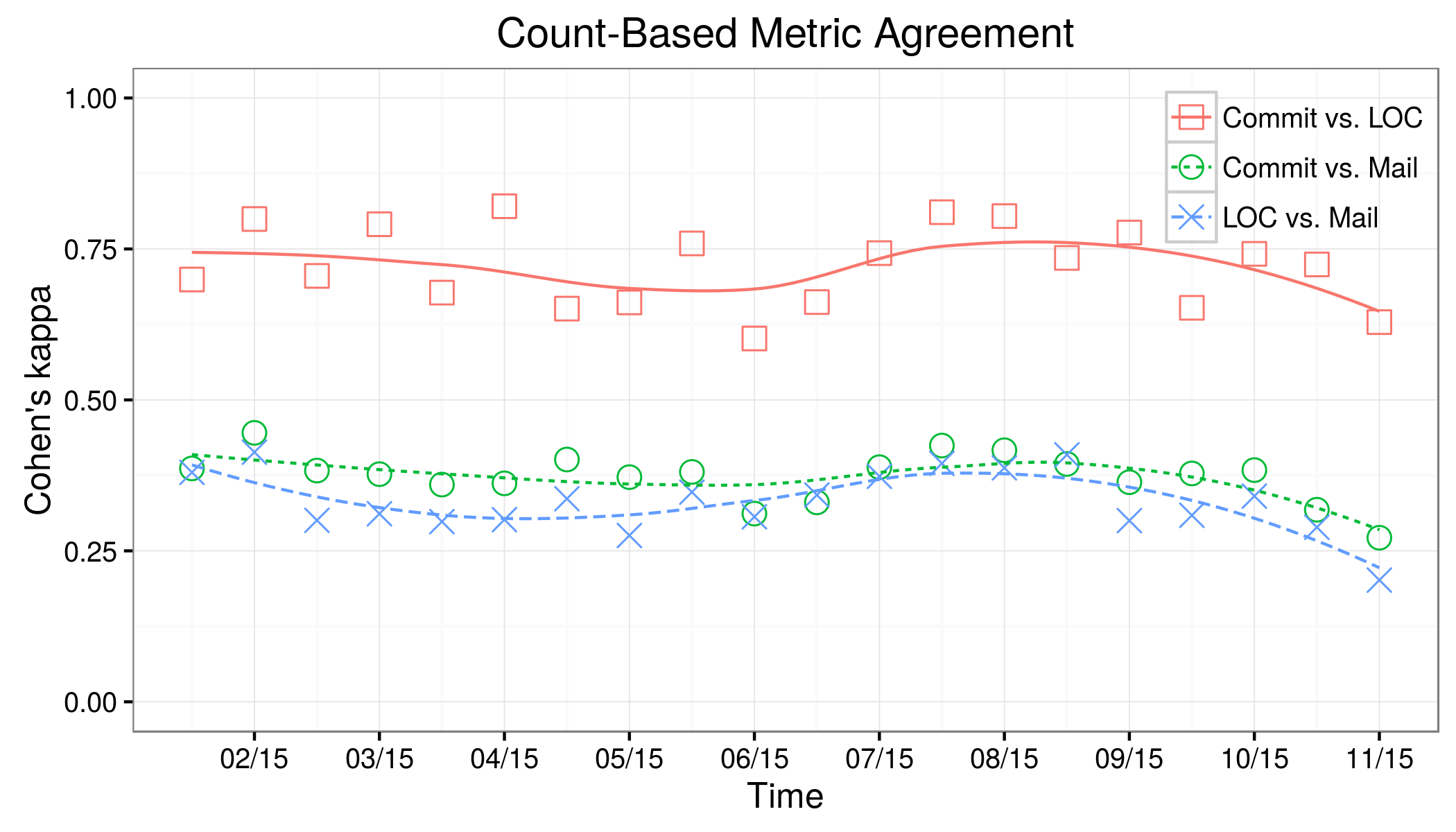}
\caption{QEMU time series representation of pairwise agreement between count-based operationalizations. The data indicate that agreement is fair to substantial and is temporally stable (i.e., mean and variance are time invariant)}.
\label{fig:agreement_ts}
\end{figure}

To address H1, we compute the pairwise agreement between all count-based metrics for a given project. For this purpose, we analyze each subject project in a time-resolved manner using a sliding-window approach (see Section~\ref{sec:methodology}) to generate time-series data that reflect the agreement for a particular three-month development period. An example time series is shown in Figure~\ref{fig:agreement_ts} for QEMU. While being only one project, the insights are consistent with the results from the other projects. The figure illustrates the agreement for Cohen's kappa, and we see that, for all comparisons, the agreement is greater than fair (e.g., greater than 0.2), which significantly exceeds the level of agreement expected by chance (see Section~\ref{sec:hypotheses}). This is evidence that the different count-based operationalizations do not contradict each other. For operationalizations that are based on the same data source (i.e., version-control system), we typically see substantial agreement (0.61--0.8). One reason for the lower cross-archive agreement could be due to problems of multiple aliases, which will be discussed in detail in Section~\ref{sec:threats}.
%Regarding Spearman's rho, we see that all metrics have high levels of correlation and are statistically different from zero (\TODO{add p-values to a table somewhere}). Typically, any value above 0.4 \TODO{add reference} is considered to be a strong correlation, and we see that, for all metrics, the value is above 0.4, indicating strong agreement in the ranking of developers' coreness. Again, we see that agreement between operationalizations based on the same archive typically exhibit higher levels of correlation than inter-archive comparisons. Interestingly, the consistently high correlation coefficients suggest that much of the disagreement seen in the dichotomous operationalizations is likely caused by forcing a dichotomous model on data that does not naturally fit into two mutually exclusive classes.
%
Another interesting result is that the agreement is relatively stable over time, which is again visible in Figure~\ref{fig:agreement_ts} for QEMU. More specifically, the arithmetic mean and variance do not significantly change over time---a property referred to as ``wide-sense stationary'' in the time-series analysis literature~\cite{hamilton1994time}. This feature of the data is a testament to the validity of the operationalizations, as we would not expect the agreement between operationalizations to change drastically from one development window to the next. The wide-sense stationary property is also important because it permits us to aggregate the data by averaging over the time windows to attenuate noise and generate more concise overviews without sacrificing scientific rigor or interpretability of the result.
\vskip 1ex
\noindent\fbox{\parbox{\linewidth}{\ Overall, the results demonstrate that the count-based operationalizations largely produce consistent results regarding the classification of developers into core and peripheral groups. We therefore \emph{accept H1}.}}

\subsection{Core and Peripheral Developers in\\Developer Networks}
We now present the manifestations of core--peripheral roles in terms of hierarchy and network positions, which are based on on structural features and role stability, which is based on structural evolution (see Section~\ref{sec:proposed_metrics}).

\begin{figure}[!t]
\centering
\includegraphics[width=0.9\linewidth]{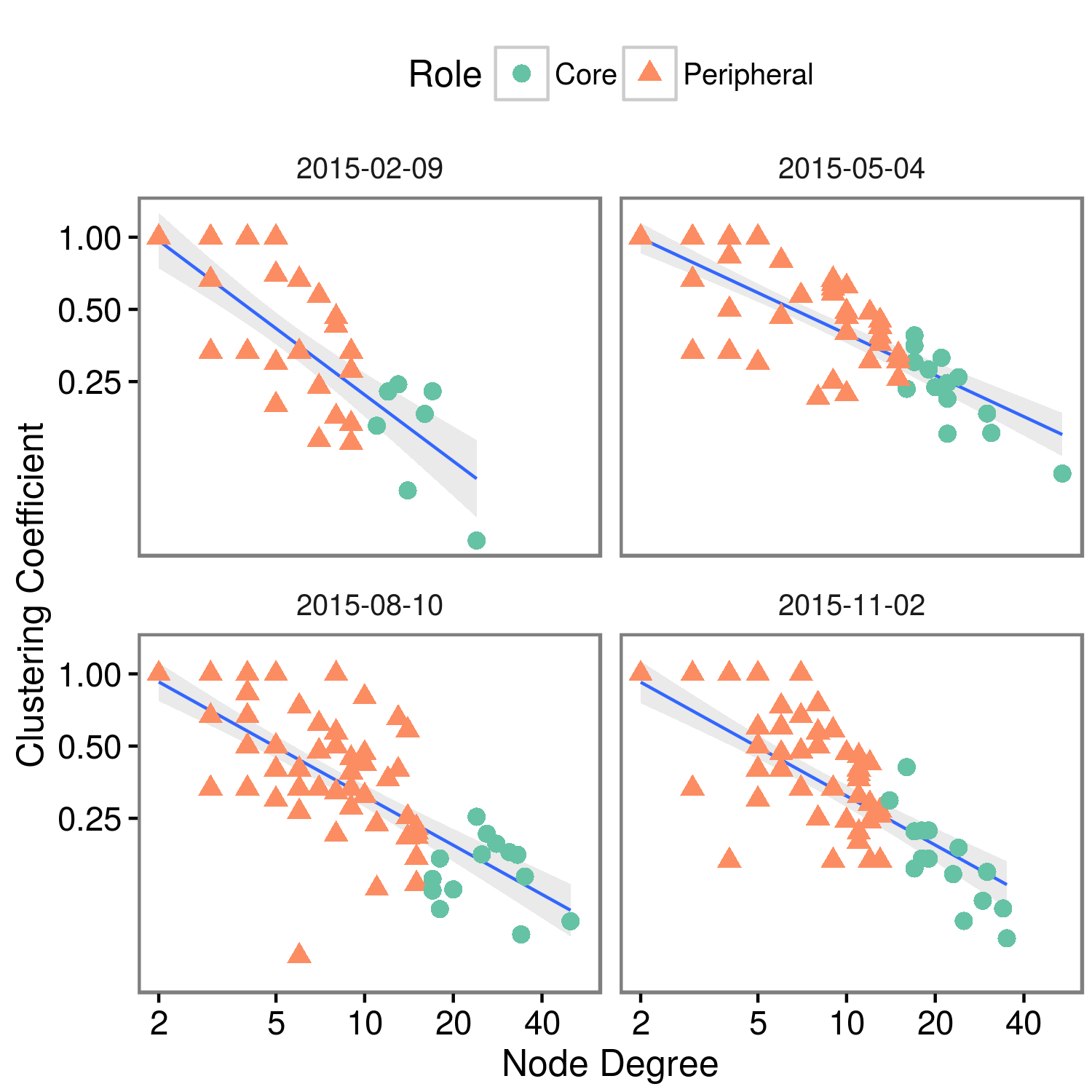}
\caption{QEMU hierarchy during four development periods. The linear dependence between clustering coefficient and degree expresses the hierarchy. Core developers should appear clustered at the top of the hierarchy (bottom right region) and peripheral developers at the bottom of the hierarchy (upper left region)}
\label{fig:dev_hier}
\end{figure}

\paragraph*{Hierarchy}
In a hierarchical network, nodes at the top of the hierarchy have a high degree and low clustering coefficient; nodes at the bottom of the hierarchy have a low degree and high clustering coefficient~\cite{ravasz2003hierarchical}. If hierarchy exists in a developer network, we should see mutual dependence between the clustering coefficient and the degree of nodes in the network~\cite{ravasz2003hierarchical}. The hierarchical relationship for QEMU is shown in Figure~\ref{fig:dev_hier}; there is an obvious dependence between the node degree and clustering coefficient. Nodes with a high degree are seen to exclusively have very low clustering coefficient and are indicative of core developers according to Section~\ref{sec:proposed_metrics}; low degree nodes have consistently higher clustering coefficients and are indicative of peripheral developers. For the remaining projects, the scatter plots are available on the supplementary Web site; here we illustrate the relationship in a more compact form in terms of Spearman's correlation coefficient between clustering coefficient and degree (see Table~\ref{table:oss_project_data} Hierarchy). We see that, for all projects, there is a strong negative correlation, indicating that the developers are indeed arranged hierarchically. In Sections~\ref{sec:network_vs_count} and \ref{sec:agreement_dev_perception}, we will see if a developer's position in the hierarchy is an organizational manifestation of their particular role.

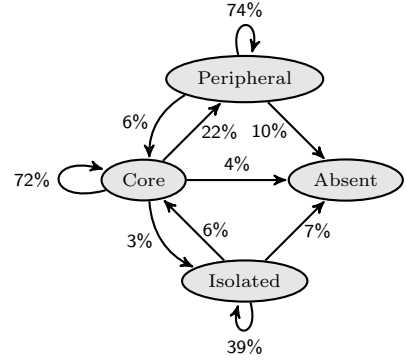
\begin{figure}[t]
\centering
\begin{tikzpicture}[->,>=stealth',shorten >=0.5pt,auto,node distance=1.9cm,
  thick,main node/.style={ellipse,fill=gray!20,draw,font=\scriptsize}]

  \node[main node] (1) {Peripheral};
  \node[main node] (2) [below left of=1] {Core};
  \node[main node] (3) [below right of=2] {Isolated};
  \node[main node] (4) [below right of=1] {Absent};

  \path[every node/.style={font=\sffamily\scriptsize}]
    (1) edge node [left] {10\%} (4)
        edge [bend right] node[left] {6\%} (2)
        edge [loop above] node {74\%} (1)
    (2) edge node [right] {22\%} (1)
        edge node {4\%} (4)
        edge [loop left] node {72\%} (2)
        edge [bend right] node[left] {3\%} (3)
    (3) edge node [right] {6\%} (2)
        edge node [right] {7\%} (4)
        edge [loop below] node {39\%} (3);
\end{tikzpicture}
\caption{The developer-group stability for QEMU shown in the form of a
Markov Chain. A few less important edges have been omitted for visual clarity.}
\label{fig:markov_chain} 
\end{figure}

\paragraph*{Stability}
Developers who fulfill a particular role within a project and who maintain their role over subsequent development periods are defined to be stable (see Section~\ref{sec:proposed_metrics}). We study this characteristic by examining the developers' transitions from one state to another (e.g., core to peripheral) in a time-resolved manner. The result of examining the developer transitions over one year of development for QEMU are shown in Figure~\ref{fig:markov_chain}. In this figure, the transition probabilities between developer states are shown in the form of a Markov chain. The primary observations are that developers in a core state are more likely to maintain their state and are substantially less likely to transition to the absent state (i.e., leave the project) or isolated state (i.e., have no neighbors in the developer network by working exclusively on isolated tasks), in comparison to developers in a peripheral state. Based on this result, the core developers represent a more stable group than peripheral developers.

\paragraph*{Core--periphery block model}
The core--periphery block model describes the core and peripheral groups, formalized as positions in a network, as a particular two-class partition of nodes (see Section~\ref{sec:proposed_metrics}). To test whether our empirical data are described by the core--periphery block model, we must compute the edge-presence probabilities for core--core, core--peripheral, and peripheral--peripheral edges. If the edge-presence probabilities are arranged according to, $p_{\text{core--core}} > p_{\text{core--periph}} > p_{\text{periph--periph}}$, then we can conclude that core developers constitute the most coordinated developers in the project, peripheral developer coordinate primarily with core developers, and peripheral developers rarely coordinate with other peripheral developers. This provides an example of a relational perspective that captures intra- and inter-relational information~(see Section~\ref{sec:proposed_metrics}). 

The edge-presence probabilities for all projects are shown in Table~\ref{table:oss_project_data} (column Edge Probabilities). In all projects, the inequality holds, indicating that the model plausibly describes our projects. The edge-presence probability for core--core has a mean value of ${4.02\times10^{-1}}$, for core--peripheral edges it is significantly lower with a mean value of ${3.30\times10^{-2}}$, and the peripheral--peripheral edge probability is lower yet with a mean value of ${1.28\times10^{-2}}$. The interpretation is that peripheral developers are twice as likely to coordinate with core developers as opposed to other peripheral developers.

Two projects are noteworthy outliers, but are still described by the core--periphery block model: Linux and PostgreSQL. For Linux, the edge-presence probabilities are notably lower in all cases, and the difference in scale between core--core edge probabilities and the others is two orders of magnitude. In the case of PostgresSQL, we see an outlier in the opposite direction. The core--core edge probability is 1, notably higher than for all other projects, much like core--peripheral edges. It is interesting that both of these projects are also outliers in terms of the size of the developer community: Linux is much larger than most projects (1510 developers), PostgreSQL is much smaller (18 developers). From this result, it appears that the scale of a project influences how likely it is for developers to coordinate, and this influence has a greater effect on the coordination of peripheral developers.
\vskip 1ex
\noindent\fbox{\parbox{\linewidth}{\ Overall, the network-based operationalizations illustrate clear manifestations of core and peripheral developer roles that agree with the abstract characteristics established by earlier empirical work. We also found evidence in terms of the core--peripheral block model that developer roles imply specific coordination preferences. On the basis of these results, we \emph{accept H2 and H3}.}}
\subsection{Agreement: Network-Based vs.\ \\ Count-Based} \label{sec:network_vs_count}
\begin{figure}[t!]
\centering
\includegraphics[width=0.8\linewidth]{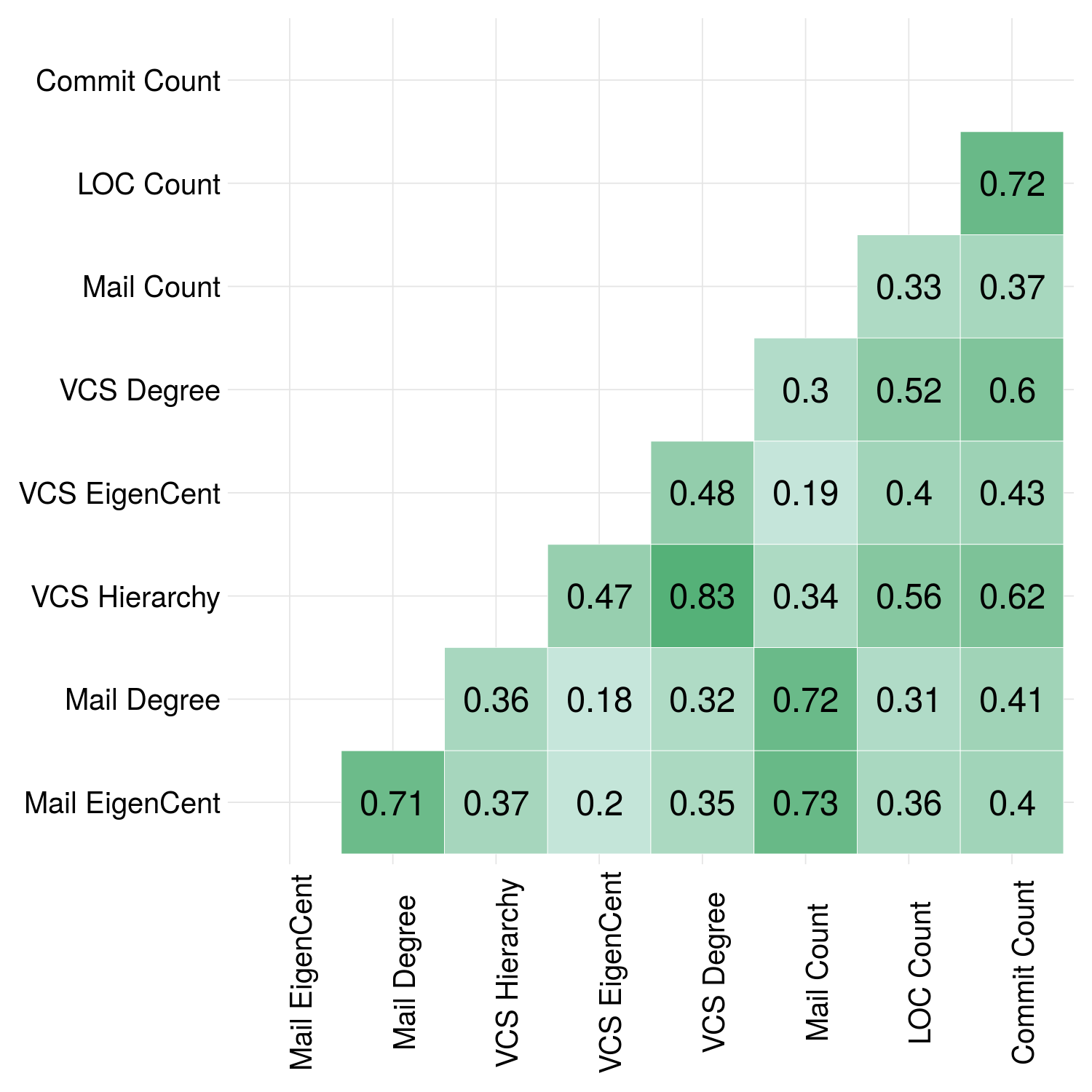}
\caption{Time-averaged agreement in terms of Cohen's kappa for QEMU. The pairwise agreement is shown for the count-based and network-based operationalizations}
\label{fig:matrix_match}
\end{figure}

So far, our results have provided evidence that the count-based operationalizations produce consistent classifications of developers, which is a testament to their validity, and that developer networks exhibit specific characteristics that are indicative of core and peripheral developer roles. Next, we present the results to relate the network-based to the count-based operationalizations for identifying core and peripheral developers. We approach this evaluation again using Cohen's kappa by averaging the level of agreement over one year of development. QEMU is used as an example project and the pairwise agreement for each operationalization is illustrated in Figure~\ref{fig:matrix_match}. The stability and core--periphery block-model operationalizations do not show up explicitly since they are derived from degree centrality.

In general, the level of agreement always exceeds 0, which indicates that the strength of agreement between all operationalizations significantly exceeds what is expected by chance. The rows/columns beginning with ``VCS'' are based on data stemming from the version-control system, and those with ``Mail'' are based on the mailing list. We again see that agreement between operationalizations defined on same data source typically have substantial agreement (0.6--0.8).
%
%The ordinal agreement, evaluated based on Spearman's rho, indicates a moderate (0.25) to very strong (\textgreater 0.75) dependence between all metrics. Similar to the dichotomous agreement, the network-based metrics generally have stronger agreement with other metrics stemming from the version control system. Based on the high values of correlation, it is highly improbable that the metrics are related only by random correspondence and that it is far more likely that they are measuring the same abstract concept. 
\vskip 1ex
\noindent\fbox{\parbox{\linewidth}{\ Overall, the results indicate that the network-based and count-based operationalizations produce classifications that are consistent. While the agreement is not perfect, the results show that the divergence from perfect agreement is similar what is seen among the count-based operationalizations. We therefore \emph{accept H4}.}}
\subsection{Agreement: Developer Perception vs.\ \\ Network-Based and Count-Based} \label{sec:agreement_dev_perception}
To establish a ground-truth classification of developers based on the perception of our survey participants, we computed the number of core and peripheral votes for each developer from the survey responses (see Section~\ref{sec:dev_survey}). For each developer, we chose the role with the highest number of votes as the ground truth and, if the count was equal, the developer was removed. Upon inspection of the responses, we found that they were largely consistent regarding a given developer's role. %To establish the ground-truth ordinal ranking, we computed a weighted average of responses for each developer.
The results of comparing the operationalizations to the ground-truth classification are shown in Table~\ref{table:survey_agreement}. Agreement was computed for 163 ground-truth samples provided by a total of 166 survey participants.\footnote{The survey response data are available at the supplementary Web site.}

The nominal agreement values, in terms of Cohen's kappa, exceed 0 indicating that all operationalizations agree with developer perception significantly more than what is expected by chance (see Section~\ref{sec:hypotheses}). The lowest agreement is seen for the count-based version-control-system metrics (i.e.,~Commit and LOC count). In contrast, all network-based operationalizations agree better (albeit in some cases only slightly) with developer perception than the basic version-control-system count-based metrics. Focusing on the comparison between different data archives, the agreement for the mailing lists metrics have even greater agreement than the corresponding version-control-system metric.
\vskip 1ex
\noindent\fbox{\parbox{\linewidth}{\ In general, the mailing list appears to more easily capture characteristics that reflect developer perception of roles. However, in many projects communication archives are not available, and in this case a network perspective on version-control system data can closely resemble the insights (regarding developer roles) provided by the communication archive. Overall, we see that a network perspective always improves the agreement with developer perception over the simpler count-based operationalizations. To this end, \emph{we accept H5}.}}
\begin{table}[t!]
\caption{Agreement with developer perception}
\vspace{1ex}
\small
\begin{center}
\label{table:survey_agreement}
\begin{tabular}{lrr}
  \toprule
 & Cohen's kappa & P value\\ 
  \midrule
  Commit Count & 0.387 & 3.12e-06 \\
  LOC Count & 0.355 & 1.91e-05 \\
  \midrule
  VCS Degree & 0.465 & 4.48e-08 \\
  VCS Hierarchy & 0.437 & 2.22e-07 \\ 
  VCS EigenCent & 0.404 & 1.74e-06 \\
  \midrule  
  Mail Count & 0.421 & 2.08e-05 \\
  Mail Degree & 0.497 & 8.23e-07 \\ 
  Mail EigenCent & 0.427 & 1.26e-05 \\ 
   \bottomrule
\end{tabular}
\end{center}
\end{table}
\subsection{Support for Relational Perspective}
\label{sec:support_network_perspective}
In addition to providing data for testing our hypotheses, the developer survey provides additional evidence for and insights into the usefulness of a relational perspective on developer roles. Our survey results suggest that developer roles are often defined in terms of differences in the mode of interaction between developers. For example, one developer wrote ``core maintainers participate in discussions on areas outside the ones that they maintain''. Only a relational perspective is able to capture this view, for example, in terms of core developers having a higher degree than peripheral developers, because they interact with developers working in areas that are distinct from the ones that they maintain. In the same vein, core developers are likely to occupy upper positions in a hierarchy, as they provide coordination bridges between the peripheral developers that have a comparatively narrow focus. Another core developer mentioned, ``I may not be contributing as much as I did in past years, but I am still active and available to answer questions from and provide guidance to other developers.'' Again, the developer has emphasized their role based on a mode of interaction with other developers. Another survey participant commented: ``The Wine project has lots of committers and a very loose structure. It's very hard to know who does what.'' A relational view on the global organizational structure has practical value to support this kind of developer awareness that is currently missing. Beside static network properties, we argue that a temporal dimension is needed to accurately operationalize developers roles, which is also supported by survey responses: ``The boundaries are fuzzy and can change over time --- sometimes I'm a core developer on libvirt, while at the present I'm only a peripheral developer'' or ``I tend to classify contributors as regular opposed to occasional.''  This is especially important as count-based operationalizations do not capture temporal relationships.

\section{Threats to Validity} \label{sec:threats}
\paragraph*{Construct Validity}
Quantifying the extent to which the operationalizations of developer roles represent the real world is one of the primary contributions of this work. We used the concept of mutual agreement as a testament to the validity of the operationalizations, however, one explanation for observing mutual agreement could be that all the operationalizations consistently reach the same wrong conclusion. While this would be a rather improbable explanation, we carried out a developer survey to provide additional evidence for that the operationalizations are valid.

For the network-based operationalizations, we used developer networks and network-analysis techniques to establish a relational basis for core and peripheral developers. This poses the threat that the networks and metrics do not accurately capture reality. This threat is minor as there is already evidence indicating that both the networks and the metrics are authentic in reflecting developer perception~\cite{joblin2015,Meneely2011}. One concern we have is regarding the unification of developers contributions, across multiple archives (i.e., mailing list and version-control system), to a single alias. However, core developers have an interest in being recognized for each contribution they make, therefore, maintaining multiple aliases would not be productive. For this reason, we think this issue has limited influence on developer classifications.

\paragraph*{Internal Validity}
We quantify the agreement between different operationalizations in terms of Cohen's kappa. For these experimental conditions, we required a probabilistic definition of agreement, because a non-error-tolerant agreement metric would be too strict to yield practical results. Cohen's kappa requires some degree of interpretation though, so we have conservatively chosen thresholds that have been established in the literature.

The results of the developer survey depend partially on individual perceptions. To limit this threat, we designed the questionnaire such that multiple developers classified the same developer and we then took the average classification to limit individual bias. 

\paragraph*{External Validity}
The results of our study are based on the analysis of 10 open-source projects. Although, the projects do represent a broad spectrum in several dimensions, they are still limited to relatively successful, mature, and large projects. Nevertheless, the results may not be relevant to immature or very small projects. Likewise, some projects, while having significant commercial involvement (e.g., Linux), are still in the end open-source and it is not yet clear if these results hold for commercial projects.

\section{Conclusion}

Software developers can play different roles in software projects. Information on 
these roles is crucial to understanding the collaborative dynamics of software projects. In particular, knowing the role of a developer provides insight regarding from whom do they likely need support or to whom could they offer support, given their current skill set and knowledge. In large, globally-distributed projects, this kind of insight can provide enormous benefits by reducing the overhead associated with developer coordination~\cite{Cataldo2010,deSouza2011}.

In an empirical study of 10 substantial open-source projects, we established evidence that commonly used count-based operationalizations of developer roles reach consistent conclusions. In particular, we found that the pairwise agreement between the operationalizations is significant and especially high when comparing operationalizations based on the same archive type. Furthermore, the agreement is temporally stable over time, which is a further testament to its validity.

Nevertheless, while offering some utility for identifying developer roles, the insights count-based operationalizations can provide are clearly limited, in particular, with regard to the manifold relationships between developers, which may even vary over time. As a novel contribution, we use developer networks to establish a relational perspective on developer roles. A key hypothesis is that developer roles should manifest distinctly in the organizational structure, which is also supported by a survey among developers. To this end, we have proposed a number of corresponding network metrics, such as positional stability, hierarchy, and a core--peripheral block model, to explore structural characteristics that capture differences between core and peripheral developers. Analyzing our 10 subject projects, we found that the network-based operationalizations largely agree with the count-based operationalizations.

While both the count-based and network-based operationalizations of developer roles hold face validity, it has not yet been shown to what extent they reflect developer perception.  Based on a survey among 166 developers, we established a ground-truth classification to address this open question. We found that all operationalizations agree with developer perception, but some align more closely than others. In particular, we found that, for count-based operationalizations, mailing-list data are more accurate in representing developer perception of roles than the version-control system. Regarding network-based operationalizations, we found that using a network perspective always increases the agreement with developer perception. 

Furthermore, our study of the temporal dimension revealed a distinction between core and peripheral developers, which is again consistent with real-world interpretations. We find this to be an important result because the count-based operationalizations do not capture temporal relationships. For example, a developer making 100 commits in one week, will appear to be equal to a developer making 2 commits per week for 50 weeks in a row, provided that the analysis window is sufficiently large.

Our results suggest that a network perspective can offer valuable insights regarding developer roles that are concealed by non-relational operationalizations. For example, the core group is comprised of the most heavily coordinated developers, and peripheral developers are more likely to coordinate with core developers than with other peripheral developers. We also found that core developers are relatively stable in the organizational structure, whereas peripheral developers tend to be more volatile. The richness of a network perspective has only begun to be explored, and we hope that our work provides the inspiration to explore further.
\pagebreak
\bibliographystyle{abbrv}
\bibliography{core_peripheral}  % sigproc.bib is the name of the Bibliography in this case
% ACM needs 'a single self-contained file'!
%\balancecolumns

\end{document}